\documentclass[
    ,final            
  ]
  {aipproc}

\layoutstyle{6x9}

\begin{document}

\title{So you want to be a lattice theorist?}


\author{Michael Creutz}{
  address={Physics Department, Brookhaven National Laboratory, Upton,
NY 11973, USA}
}

\begin{abstract}
For this after dinner talk I intersperse images of real lattices with
a discussion of the motivations for lattice gauge theory and some
current unresolved issues.
\end{abstract}

\maketitle


\begin{figure}
  \includegraphics[height=.38\textheight]{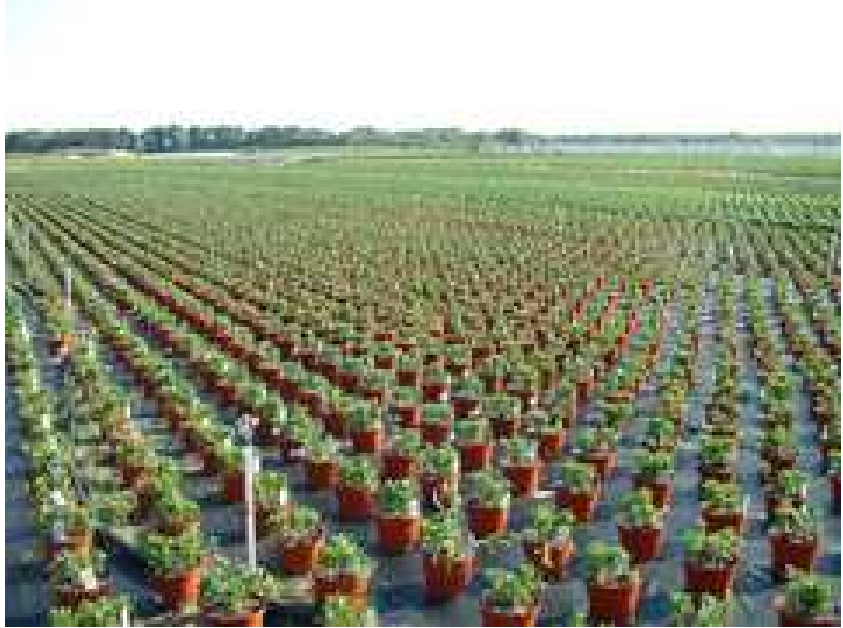}
  \caption{Lattices abound in the real world.  Here is lattice of
  daisies.}
  \label{daisylattice}
\end{figure}

Although lattices are frequently seen in the real world, as in Figure
\ref{daisylattice}, to the particle theorist they are nothing but a
mathematical trick.  We constrain quarks so that rather than following
arbitrary world lines, they only move in discrete hops between lattice
sites.  As they hop they get spun around in group space by the gauge
fields, which are restricted to the lattice bonds.  It is a nice
framework for exploring confinement, which is related to this
spinning; quarks act like kangaroos, strongly preferring to hop
together in mobs.

Since the vacuum is not a crystal, this seems at first sight a rather
strange thing to do.  However, the lattice has several advantages,
primarily in allowing calculations in situations where other methods
fail.  In particular, one can go far beyond the realms of perturbation
theory or semi-classical methods.  Furthermore, the predictions can
have crucial experimental implications.  These extend to many areas of
particle and nuclear physics, from extracting weak matrix elements in
processes involving large hadronic corrections, to understanding the
behavior of matter under the extreme conditions of heavy ion
collisions, and to detailed studies of hadronic structure.

\begin{figure}
  \includegraphics[height=.3\textheight]{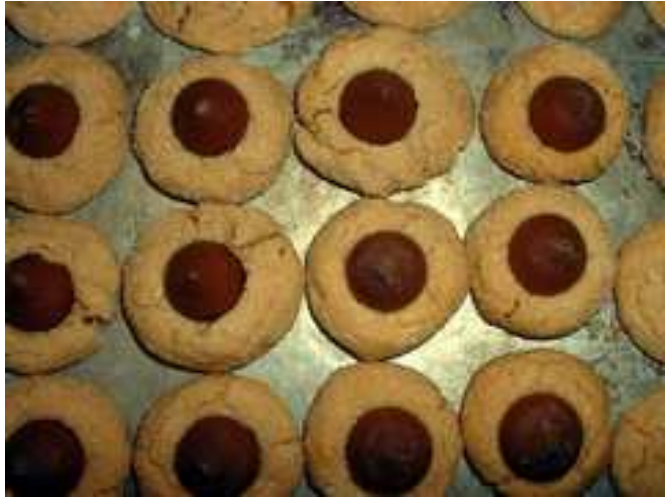}
  \caption{Lattices can have good flavors.  But beware of lurking
    tastes.}
  \label{cookies}
\end{figure}

\begin{figure}
  \includegraphics[height=.3\textheight]{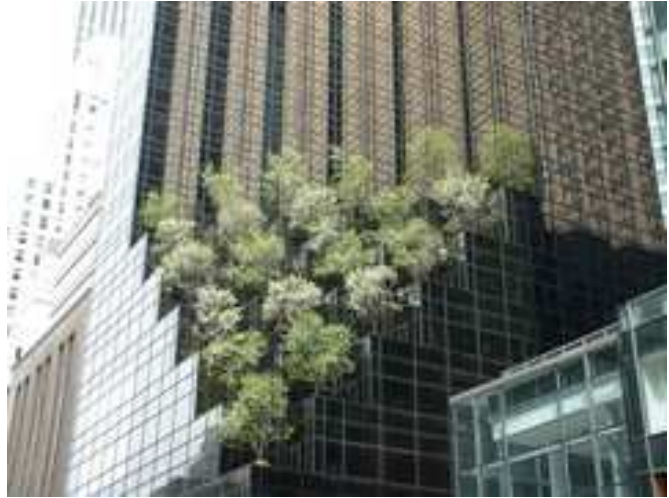}
  \caption{A lattice of trees surrounded by other lattices in New York
  City.}
  \label{trees}
\end{figure}

And of course we get to have fun playing with big computers.  Indeed,
these themselves are large lattices of processors, such as the six
dimensional torus that makes up the QCDOC supercomputer dedicated to
lattice gauge theory.  There are also more abstract reasons to study
lattice gauge theory.  As shown in Figure \ref{cookies}, lattices can
have good flavors.  However one should be careful of any harmful
lurking tastes.  Lattices are frequently seen in cities, such as the
lattice of trees seen in Figure \ref{trees}.

One of the fun things about lattice gauge theory is the addictive
power it gives over the system.  Entire lattice configurations are
stored in the computer memory, and you are free to measure anything
you want.  In the process uncertainties can arise, and the theorist is
in the unusual situation of having error bars.  First of all, since we
are using Monte Carlo methods, there will be statistical errors.
These can be reduced by massive applications of computer time.  There
are also several sources of systematic error, some of which we have
control over.  These include finite volume and finite lattice spacing
corrections, which can also be reduced by increased computer time.  In
practice using quarks with physical masses is quite computer
intensive; so, we usually simulate with heavier than normal quarks and
then do an extrapolation.

There are also some sources of error that are basically uncontrolled.
One is the so-called ``valence'' or ``quenched'' approximation,
wherein the feedback of internal quark loops is ignored.  This is a
tempting approximation since it saves a couple of orders of magnitude
of computer time.  But fortunately the continuing growth in computer
power is now alleviating the need for this inexact approach.

Another uncontrolled source of error comes from extrapolations in the
number of quark flavors.  Again to save computer time, it is popular,
mainly in the US, to start with a fermion formulation that has some of
the naive doubling issues remaining and then do an extrapolation down
to the desired number of quark species.  This is done by replacing the
fermion determinant by a non integer power.  Since the starting
determinant is not a power, this procedure has not been theoretically
justified.  Indeed, it explicitly gives incorrect behavior in the
chiral limit of small masses.  I will return to this issue later.

\begin{figure}
  \includegraphics[height=.3\textheight]{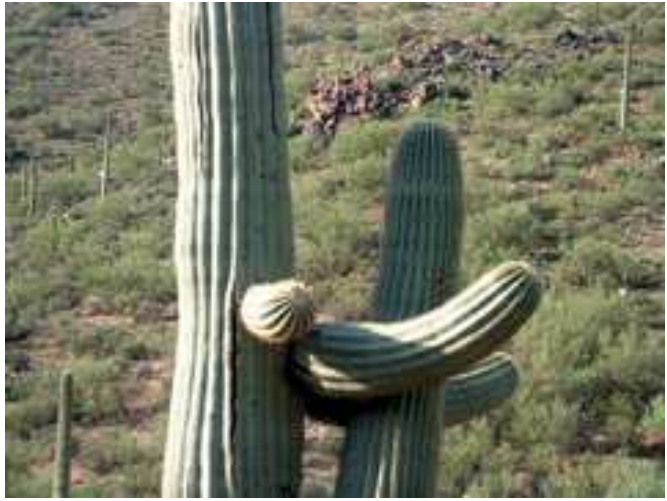}
  \caption{After recent advances, the lattice now embraces
  chiral symmetry.}
  \label{embrace}
\end{figure}

Sometimes the lattice can reveal rather subtle issues.  In particular,
for many years the way chiral symmetry worked on the lattice was
puzzling.  We know chiral symmetry is important to the lightness of
the pion, which is theoretically tied to the lightness of the up and
down quarks.  The lattice removes all infinities, and thus issues such
as anomalies coming from divergences can be tricky.  Ignoring these
anomalies forces the theory to cancel them with extra species, known
as doublers.  But recent years have seen the development of elegant
approaches that solved these problems.  One tack considers our four
dimensional world as an interface in five dimensions
\cite{Kaplan:1992bt,Furman:1994ky}.  An alternative extracts the
essence of this interface into the slightly non-local overlap operator
\cite{Neuberger:1997fp}.  This satisfies an elegant modification of
naive chiral symmetry.  So, as indicated in Figure \ref{embrace}, the
lattice and chiral symmetry now get along nicely.

Despite these advances, there remain some subtle unsolved problems in
lattice gauge theory.  One of these involves the standard model, where
the weak gauge fields are coupled in a parity violating manner.
Neutrinos are experimentally known to spin only to the left, but all
known lattice formulations also bring in right handed partners.  For
example, with domain wall fermions there is naturally present an
anti-wall which couples with equal strength to the gauge fields.  Ad
hoc Higgs fields can give the mirror particles a different mass, but
they are always there.  To the extend that the lattice is a technique
to define a field theory, this raises worries that the usual standard
model might be incomplete or even not well defined.

\begin{figure}[b]
  \includegraphics[height=.25\textheight]{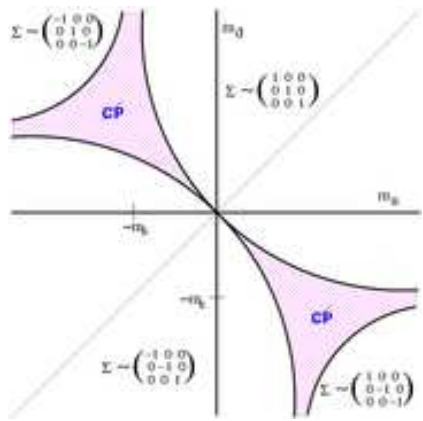}
  \caption{The phase structure expected for three flavor QCD as the up
  and down quark masses are varied at fixed strange quark mass.
  Spontaneous CP violation occurs in regions where the up and down
  quark masses differ in sign.  No structure appears when just a
  single quark mass vanishes.}
  \label{threeflavor}
\end{figure}

The other major unsolved problem involves the properties of matter at
high baryon density.  Here there are no practical known algorithms for
simulations.  Monte Carlo methods fail because there is no positive
measure for the path integral.  All existing attempts to circumvent
this issue require computer time growing exponentially with the system
size.  This is particularly frustrating in light of the rich phase
diagram expected at high density, filled with exotic phenomena such as
color superconductivity.

There are some lattice topics which are highly controversial.  I will
illustrate the issue starting from a conventional continuum discussion
of how chiral symmetry works in three flavor QCD.  Here a longstanding
tool comes from effective chiral Lagrangians.  The physics of the
light pseudoscalars is nicely modeled in terms of an effective field
$\Sigma$ which lies in the group $SU(3)$.  Incorporating quark masses
into this picture involves a potential of the form $V(\Sigma)=-{\rm
Tr}\ M\Sigma$, where the mass matrix is
\begin{equation}
M=\pmatrix{m_u & 0 & 0\cr
           0 & m_d & 0\cr
           0 & 0 & m_s\cr
}
\end{equation}
As we vary the quark masses, minimizing this potential predicts a rich
phase structure \cite{Creutz:2003xu}, sketched in Figure
\ref{threeflavor}.  Indeed, I discussed this structure at length
during the previous meeting in this series \cite{Creutz:2004zg}.
Striking features are the regions of spontaneous CP violation where
the minima of the potential are doubly degenerate at complex values of
$\Sigma$.

\begin{figure}
  \includegraphics[height=.3\textheight]{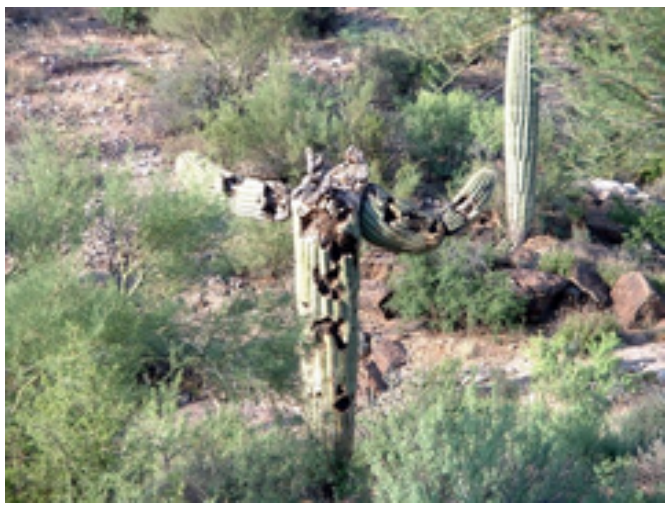}
  \caption{Controversial ideas came to the front at Lattice 2006.}
  \label{shootout}
\end{figure}

An important feature of this diagram is the absence of any special
features when only a single quark mass vanishes.  The presence of the
other quark masses is sufficient to stabilize the vacuum value for
$\langle\Sigma\rangle$, which is real and not accompanied by any exact
massless modes.  This is a consequence of the anomaly at work;
massless Goldstone particles require more than one quark mass to
vanish at the same time.

The controversy concerns a numerical algorithm that is incapable of
seeing this structure.  The feelings here are rather strong, as shown
in Figure \ref{shootout} from Lattice 2006.  The ``staggered cabal''
promotes using a technique known as ``rooted staggered quarks.''  This
is the procedure mentioned above of starting with extra particles and
taking a fractional power of the fermion determinant.  The issue that
arises is that the starting staggered formulation has an exact chiral
symmetry when any single quark mass vanishes.  This symmetry survives
the rooting process, and demands the existence of a massless Goldstone
mode where the simple effective chiral Lagrangian says there is none.
Indeed, this is in direct contradiction with known anomalies
\cite{Creutz:2006ys}.

The condoners of this algorithm \cite{Bernard:2006vv} suggest, without
proof, that these evils will drop away in the continuum limit as long
as one avoids the zero quark mass axes in Figure \ref{threeflavor}.
They argue that there is actually a plethora of extra particles, one
of which is this unwanted Goldstone mode, but their total contribution
cancels as the continuum limit is taken.  For three flavors using
independent rooted staggered quarks, there are 144 pseudoscalar
bosons, out of which only the usual 9 should survive the continuum
limit.  This requires a loss of unitarity so that the total cross
sections to produce some of these extra particles can be negative.
Also the extra massless particle induces long range forces that make
the algorithm non-local.  And all of these unproven conjectures are
being made just to save some computer time over other algorithms, such
as Wilson, domain wall, or overlap fermions, that do not so severely
mutilate the qualitative chiral behavior, I conclude that rooting can
be unhealthy, although the extreme contortions being tried to rescue
the approach might be amusing enough to warrant a movie.

\begin{figure}
  \includegraphics[height=.35\textheight]{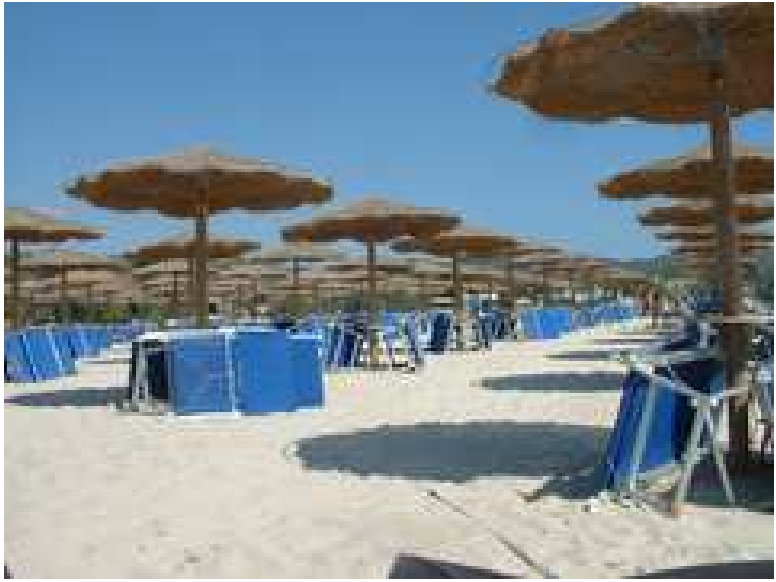}
  \caption{A lattice of palapas at the 2004 meeting in this series,
  held in Villasimius, Sardinia.}
  \label{sardinia}
\end{figure}

I conclude with one final reason one might want to be a lattice
theorist.  We often meet in very nice places to search out new
lattices, such as the marble/basalt arrays here in the Azores or the
environment shown in Figure \ref{sardinia} from the 2004 meeting in
this series.  And of course, as you will see tomorrow night, this
meeting has a strong tradition of taking poster sessions seriously!

\begin{theacknowledgments}
This manuscript has been authored under contract number
DE-AC02-98CH10886 with the U.S.~Department of Energy.  Accordingly,
the U.S. Government retains a non-exclusive, royalty-free license to
publish or reproduce the published form of this contribution, or allow
others to do so, for U.S.~Government purposes.
\end{theacknowledgments}

\end{document}